\documentclass[numberedappendix]{emulateapj}
\usepackage{apjfonts,natbib}
\defcitealias{2001PASJ...53....1K}{K01}

\newcommand{\cs}{c_\mathrm{s}}
\newcommand{\oring}{\omega_\mathrm{ring}}
\newcommand{\ospiral}{\omega_\mathrm{spiral}}
\newcommand{\OK}{\Omega_\mathrm{K}}
\newcommand{\rISCO}{r_\mathrm{ISCO}}
\newcommand{\rpeak}{r_\mathrm{peak}}
\newcommand{\kmax}{\kappa_\mathrm{max}}

\begin{document}

\title{Oscillations of the Inner Regions of Viscous Accretion Disks}

\author{Chi-kwan Chan}

\affil{Institute for Theory and Computation,
  Harvard-Smithsonian Center for Astrophysics,
  60 Garden Street, Cambridge, MA 02138}

\email{ckchan@cfa.harvard.edu}

\begin{abstract}
  Although quasi-periodic oscillations (QPOs) have been discovered in
  different X-ray sources, their origin is still a matter of debate.
  Analytical studies of hydrodynamic accretion disks have shown three
  types of trapped global modes with properties that appear to agree
  with the observations.  However, these studies take only linear
  effects into account and do not address the issues of mode
  excitation and decay.  Moreover, observations suggest that
  resonances between modes play a crucial role.  A systematic,
  numerical study of this problem is therefore needed.  In this paper,
  we use a pseudo-spectral algorithm to perform a parameter study of
  the inner regions of hydrodynamic disks.  By assuming
  $\alpha$-viscosity, we show that steady state solutions rarely
  exist.  The inner edges of the disks oscillate and excite
  axisymmetric waves.  In addition, the flows inside the inner edges
  are sometimes unstable to non-axisymmetric perturbations.
  One-armed, or even two-armed, spirals are developed, which provides
  a plausible explanation for the high-frequency QPOs observed from
  accreting black holes.  When the Reynolds numbers are above certain
  critical values, the inner disks go through some transient turbulent
  states characterized by strong trailing spirals; while large-scale
  leading spirals developed in the outer disks.  We compared our
  numerical results with standard thin disk oscillation models.
  Although the non-axisymmetric features have their analytical
  counterparts, more careful study is needed to explain the
  axisymmetric oscillations.
\end{abstract}


\section{Introduction}
\label{sec:introduction}

Quasi-periodic oscillations (QPOs) are strong coherent features in the
power density spectra (PDS) that are found in different low mass X-ray
binaries (LMXBs, see the review article by
\citealt{2006csxs.book...39V} and \citealt{2006ARA&A..44...49R}).
Their frequencies range from $\sim 300$ to $\sim 1200$ Hz, suggesting
that they correspond to accretion flows very close to the central
objects \citep{2000ARA&A..38..717V, 2006csxs.book...39V}.
Understanding the origin of QPOs will, therefore, lead to precise
measurements of black hole spins as well as direct tests of Einstein's
equivalent principle (\citealt{2004AIPC..714...29P}, but also see
\citealt{2008PhRvL.100i1101P}).  Although several models have been
proposed in the last decade to explain some aspects of the
observations, the origin of QPOs is still a matter of debate.

Local analytical studies of hydrodynamic accretion disks have shown
three types of trapped global modes (see the review article by
\citealt{2001PASJ...53....1K}, hereafter
\citetalias{2001PASJ...53....1K}, and references therein, or more
recently \citealt{2008NewAR..51..828W}).  The coupling between
inertial oscillations and pressure variations lead to two different
frequencies.  The higher frequencies correspond to inertial acoustic
$p$-modes and the lower ones correspond to gravity $g$-modes
\citep{1997ApJ...476..589P, 2002ApJ...567.1043O}.  Oscillations in the
vertical direction are called corrugation $c$-modes
\citep{2001ApJ...548..335S}.  Although these oscillations have
properties that appear to agree with the observations, analytical
studies can take only the linear effects into account and they do not
address issues of mode excitation, propagation, and decay.  Moreover,
observations of QPOs in black holes strongly suggest that resonances
between modes play a crucial role \citep{2001A&A...374L..19A,
  2001AcPPB..32.3605K}.  This fact has led to the development of
non-linear models \citep[for example,][]{2004PASJ...56..905K,
  2008PASJ...60..111K, 2008arXiv0808.1218K}.

In principle, numerical simulations can be used to test these models,
including both linear and non-linear ones.  One would like to carry
out simulations for many dynamical time scales to obtain accurate PDS
and study their temporal variability.  However, even with the
\citet{1973A&A....24..337S} $\alpha$-viscosity (which greatly
simplifies the problem) the effective/turbulent Reynolds numbers in
thin accretion disks are still beyond the value that one could perform
\emph{direct} numerical simulations with standard numerical methods.

We have developed earlier a two-dimensional, pseudo-spectral algorithm
in order to model thin accretion flows around black holes
\citep{2005ApJ...628..353C}.  Spectral methods are high-order
numerical methods that have a very small numerical dissipation.  For
two-dimensional flows, they are at least an order of magnitude more
efficient then (low order) finite difference methods.  Our algorithm
solves the hydrodynamic equations in the disk plane (i.e., with $r$
and $\phi$) by using the thin disk approximation.  In addition to the
$\alpha$-viscosity, high-order artificial viscosity is implemented by
using a spectral filter, which affects only the large wavenumber modes
and preserves correct shock properties \citep[see][for
  details]{291525, 291526}.  We use a buffer zone to absorb outgoing
waves at the outer boundary.  For the inner boundary, because the flow
is always super-sonic (toward the central object), no explicit
boundary condition is needed.  We apply this algorithm to study the
very inner regions of two-dimensional, viscous, polytropic accretion
disks around a black hole with zero spin.

We find that steady state solutions do not exist for most combinations
of the parameters.  The inner edges of the disk often oscillate and
excite axisymmetric waves with propagates all the way to the outer
boundary.  Other than that, non-axisymmetric spirals are sometimes
developed in the simulations and provide a plausible explanation of
the QPOs.  Depending on the values of the parameters, these spirals
can be classified as inner one-armed spirals or inner two-armed
spirals; in the cases with extreme Reynolds numbers, the results are
steady global one-armed spirals.  Although detailed study of all these
features is beyond the scope of this paper, we will make connections
between the observed numerical phenomena and the standard analytical
models.

This paper is organized as following.  The assumptions and the
governing equations of the problem are presented in the next section.
In \S\ref{sec:ic}, we derive an approximate steady state solution,
which is used to initialize our simulations.  We summarize the results
from our simulations in \S\ref{sec:results} and describe our temporal
analysis method in \S\ref{sec:fft}.  Detailed discussions of different
important features are presented in the subsequent sections, namely,
the axisymmetric rings in \S\ref{sec:rings}, non-axisymmetric spirals
in \S\ref{sec:spirals}, and the steady global spirals in
\S\ref{sec:global}.  We describe the limitations of our study in
\S\ref{sec:limitations}.  Finally, we conclude with a discussion and
suggest future research direction in \S\ref{sec:discussions}.

\section{Assumptions and Equations}
\label{sec:assumptions}

We are interested in the dynamics of thin accretion disks.  For
simplicity, we neglect the motion in the vertical direction.  By
assuming a polytropic equation of state, the hydrodynamics is fully
described by only three quantities, namely, the column density
$\Sigma$ and the two velocity components in the disk plane $v_r$,
$v_\phi$.  The governing equations are thus given by the continuity
equation
\begin{equation}
  \frac{\partial\Sigma}{\partial t} + \nabla\cdot(\Sigma\mathbf v) = 0
  \label{eq:continuity}
\end{equation}
and the (two-dimensional) Navier-Stokes equation
\begin{equation}
  \Sigma\frac{\partial\mathbf{v}}{\partial t} +
  \Sigma(\mathbf{v}\cdot\nabla)\mathbf{v} = -\nabla P +
  \nabla\mathbf\tau + \Sigma\;\mathbf{g}.
  \label{eq:navier-stokes}
\end{equation}
In the above equations, $P = K \Sigma^\Gamma$ is the height-integrated
thermal pressure with $K$ and $\Gamma$ being the polytropic constant
and polytropic index.  We denote by $\tau$ the viscosity tensor, which
(in Cartesian coordinates) is given by
\begin{equation}
  \tau_{i\!j} = \nu_\mathrm{SS}\Sigma\left[\left(\frac{\partial
  v_i}{\partial x_j} + \frac{\partial v_j}{\partial x_i}\right) -
  \frac{2}{3}(\nabla\cdot\mathbf{v})\;\delta_{ij}\right].
\end{equation}
We also define the sound speed $\cs \equiv (\Gamma P / \Sigma)^{1/2}$
and the scale height $H \equiv \cs/\OK$, where $\OK$ is the
``Keplerian'' frequency defined below.

We employ the \citeauthor{1973A&A....24..337S} $\alpha$-prescription
so that the effective kinematic viscosity is parametrized by the
dimensionless number $\alpha \lesssim 1$ in the form $\nu_\mathrm{SS}
\equiv \alpha \cs H$.  The effects of general relativity is taken into
account within the framework of the pseudo-Newtonian approximation,
for which
\begin{equation}
  \mathbf g \equiv - \frac{GM}{(r - r_\mathrm{S})^2} \mathbf{\hat r},
\end{equation}
where $r_\mathrm{S} = 2GM/c^2$ is the Schwarzschild radius and
$\mathbf{\hat r}$ is the unit vector in the positive r-direction.
Therefore, the ``Keplerian'' frequency is given by
\begin{equation}
  \OK \equiv \frac{\sqrt{GM/r}}{r-r_\mathrm{S}}.
\end{equation}
We define our variables in terms of the natural length and time scales
$GM/c^2$ and $GM/c^3$.  The radial domain is chosen to be $4 \le r \le
64$ so it contains the innermost stable circular orbit (ISCO) at
$\rISCO = 6$.  We run the simulations up to $t = 20,000$, which is
about 325 orbital time scale at $\rISCO$.  All three dynamic variables
are output at every time unit, resulting into $20,001$ snapshots
(including the initial condition at $t = 0$).

There are three (physical) parameters in our equations, namely, the
polytropic constant $K$, the polytropic index $\Gamma$, and the
\citeauthor{1973A&A....24..337S} $\alpha$ parameter.  The accretion
rate $\dot{M}$, instead of being an input parameter as in
\citet{1996MNRAS.283..919M} or \citet{2008arXiv0805.0598M}, is
computed from our simulations.  We choose three values for each of the
parameters.  Based on the thin disk assumption, we use $K = 10^{-4}$,
$10^{-3.5}$, and $10^{-3}$ with $\Gamma = 1$, $4/3$, $5/3$.  We also
take $\alpha = 0.1$, $10^{-0.5}$, $1$, which provide Reynolds numbers
$Re \lesssim 10^5$.  The parameter study, therefore, contains 27
different simulations (see Table~\ref{tab:para}).

\section{Initial Conditions}
\label{sec:ic}

To avoid numerical instabilities at the inner boundary, we start the
simulations at a state in which the flow close to the inner boundary
is falling toward the central object.  Because the numerical value of
$K$ we choose is small, we neglect the pressure and viscosity term.
and solve an approximated solution of equations~(\ref{eq:continuity})
and (\ref{eq:navier-stokes}).  Using the superscript $^{(0)}$ to
denote the zeroth order quantities, the zeroth order hydrodynamic
equations reduce to the very simple form
\begin{eqnarray}
   \frac{1}{r}\frac{\partial}{\partial r}(r v_{r}^{(0)} \Sigma^{(0)}) & = & 0,
   \label{eq:continuity0} \\ 
   v_{r}^{(0)}\frac{\partial v_{r}^{(0)} }{\partial r} & = &
   \frac{v_{\phi}^{(0)\;2}}{r} - \frac{GM}{(r-r_\mathrm{S})^2}, \\
   v_{r}^{(0)}\frac{\partial v_{\phi}^{(0)}}{\partial r} & = &
   -\frac{v_{r}^{(0)}v_{\phi}^{(0)}}{r}.
\end{eqnarray}
Integrating equation~(\ref{eq:continuity0}), we obtain
\begin{equation}
  2\pi r v_{r}^{(0)} \Sigma^{(0)} = -\dot{M}. \label{eq:M_dot}
\end{equation}
The accretion $\dot{M}$ here is nothing but a constant of integration.

For the Navier-Stokes equation, there are two different classes of
solutions.  The first class is the Keplerian solution (KS):
\begin{eqnarray}
  v_{r}^{(0)} & = & 0, \\
  v_{\phi}^{(0)} & = & r\;\OK,
\end{eqnarray}
which is obtained by solving two algebraic equations; there is no
constant of integration.  The column density $\Sigma^{(0)}$ in this
solution is completely arbitrary (as long as it is non-negative) and
$\dot{M} = 0$.

The second class of solutions is the free-falling solution (FS):
\begin{eqnarray}
  e & = & \frac{v_{r}^{(0)\;2}}{2} + \frac{l^2}{2r^2} -
  \frac{GM}{r-r_\mathrm{S}}, \label{eq:e} \\
  l & = & v_{\phi}^{(0)} r, \label{eq:l}
\end{eqnarray}
which are simply the conservation laws of energy and angular
momentum. The constant of integration $e$ has the meaning of specific
energy and $l$ has the meaning of specific angular momentum,
respectively.  The column density is obtained by combining
equation~(\ref{eq:M_dot}) and (\ref{eq:e}).

For the region inside the ISCO, the rotational profile should follow
FS.  On the other hand, the disk is (almost) Keplerian far away from
the central object.  Hence, by requiring that the two classes of
solutions match each other at some critical radius $r_\mathrm{c}$
outside ISCO, the zeroth-order rotation profile is
\begin{equation}
  \Omega^{(0)} = \left\{\begin{array}{lll}
   (r_\mathrm{c}/r)^2   \OK(r_\mathrm{c}) &, & r < r_\mathrm{c} \\
  \ \ \ \ \ \ \ \ \ \ \ \OK(r) &, & r \ge r_\mathrm{c}.
  \end{array}\right.
\end{equation}
It is convenient to define
\begin{equation}
  q^{(0)} \equiv -\frac{d\ln\Omega^{(0)}}{d\ln r}
  = \left\{\begin{array}{lll} 2 &, & r < r_\mathrm{c} \\
  (3r - r_\mathrm{S})/2(r-r_\mathrm{S}) &, & r \ge r_\mathrm{c}
  \end{array}\right.
\end{equation}
as the background ``shear''.  The value of $r_\mathrm{c}$ is a free
parameter in the zeroth-order solution, although it should not be too
big compared to $\rISCO = 6GM/c^2$.

Because of the thin disk assumption, the first order azimuthal
velocity is small $v_{\phi}^{(1)}$ compared to $v_{\phi}^{(0)}$.
Instead of using the radial velocity equation, we can easily obtain
the first order radial velocity $v_{r}^{(1)}$ by integrating the
angular moment equation, which gives
\begin{equation}
  -\nu_\mathrm{SS}\Sigma^{(0)}\frac{d\Omega^{(0)}}{dr} =
  -v_{r}^{(1)}\Sigma^{(0)}\Omega^{(0)} -
  \frac{\dot{L}}{2\pi r^3}.
\end{equation}
The constant of integration, $\dot{L}$, describes the rate of angular
momentum transport.  Substituting for $\nu_\mathrm{SS}$ and recalling
the definitions of $q$ and $\dot{M}$, we obtain
\begin{equation}
  2\pi q\alpha\Gamma K\frac{\Sigma^{(0)\;\Gamma}}{\OK} =
  \dot{M} - \frac{\dot{L}}{r^2\Omega^{(0)}}.
  \label{eq:general}
\end{equation}
Note that we differentiate $\OK$, which originates from the disk scale
height (and hence vertical gravity), from the zeroth order angular
velocity $\Omega^{(0)}$.  Taking the limit $r\rightarrow\infty$ and
assuming that $\Sigma$ decays slower than $r^{-2}$ (the
\citeauthor{1973A&A....24..337S} solution gives $\Sigma \sim r^{-3/4}$
for large $r$), the above equation reduces to
\begin{equation}
  \dot{M} = 3\pi\alpha\Gamma K\lim_{r\rightarrow\infty}
  \frac{[\Sigma^{(0)}]^{\Gamma}}{\OK}.
  \label{eq:accretion_rate}
\end{equation}
We define the quantity $\mathcal{L} \equiv
\lim_{r\rightarrow\infty}\Sigma^{(0)\;\Gamma}/\OK$.  In order to have
a non-zero accretion rate, $\mathcal{L}$ should converge to some
positive value.  If the fluid is isothermal, this value simply sets
the normalization of the density and does not affect the dynamics.
When $\Gamma \ne 1$, $\mathcal{L}$ defines the sound speed and changes
the behavior of the accretion disk.

In standard accretion disk models, the value of $\dot{L}$ is usually
solved by assuming $q = 0$ at some boundary layer
\citep[see][]{2002apa..book.....F}.  However, for accreting black
holes, it is more natural to assume that the density drop close to
zero around the ISCO.  Let $\epsilon$ be a small parameter.
Equation~(\ref{eq:general}) can be rewritten as
\begin{equation}
  \dot{L} = (1-\epsilon)\dot{M}
  \frac{\rISCO^2}{\rISCO-r_\mathrm{S}}
  \sqrt{\frac{GM}{\rISCO}}.
\end{equation}
The density outside the critical radius is then given by
\begin{equation}
  q\frac{\Sigma^{(0)\;\Gamma}}{\OK} = \frac{3}{2}
  \left[1-(1-\epsilon) \left(\frac{\rISCO}{r}\right)^{3/2}
  \frac{r - r_\mathrm{S}}{\rISCO - r_\mathrm{S}}
  \right] \mathcal{L}.
  \label{eq:limit}
\end{equation}
The radial velocity can then be solved by the continuity equation
\begin{equation}
  v_{r}^{(1)} = -\frac{\dot{M}}{2\pi r \Sigma^{(0)}}.
\end{equation}

Inside the ISCO, gravity is the dominant effect.  We solve the radial
velocity by conservation of energy
\begin{eqnarray}
  v_{r}^{(1)} & = & -\left[v_{r\mathrm{ISCO}}^2 +
  l_\mathrm{ISCO}^2\left(\frac{1}{\rISCO^2} -
  \frac{1}{r^2}\right)\right. \nonumber\\ & & \ \ \ \ \ \ \left. -
  2GM\left(\frac{1}{\rISCO-r_\mathrm{S}} -
  \frac{1}{r-r_\mathrm{S}}\right)\right]^{-1/2}.
\end{eqnarray}
Here we denote by $v_{r\mathrm{ISCO}}$ and $l_\mathrm{ISCO}$ the
radial velocity and specific angular momentum at the ISCO,
respectively.  The parameter $\epsilon$ affects the above equation
through $v_{r\mathrm{ISCO}}$.  It should be chosen small enough so
that the constraint $v_{r\mathrm{ISCO}} \ll l_\mathrm{ISCO}/\rISCO$ is
always satisfied.

The approximation described here introduces two extra parameters.  The
first one is $\epsilon$, which controls the initial transport of
angular moment.  In our simulation, it is always taken $\epsilon =
1\%$.  The exact value of $\epsilon$ is not important because the
initial conditions evolve to the correct steady state solution in a
few dynamical time scales.  The second one is the limiting term
$\mathcal{L} \equiv \lim_{r\rightarrow\infty} \Sigma^{(0)\;\Gamma} /
\OK$.  It is chosen so $\Sigma^{(0)}(r_{\max}) = 1$ and $\epsilon = 0$
in equation~(\ref{eq:limit}), where $r_{\max} = 64$ is the outer
radius of our computational domain.  This makes $\Sigma =
\mathcal{O}(1)$ so the numerical values of $K$ roughly represent
$\cs^2$.

\section{Overview of Results}
\label{sec:results}

\begin{figure*}
  \includegraphics[scale=0.75,trim=18 18 0 12]{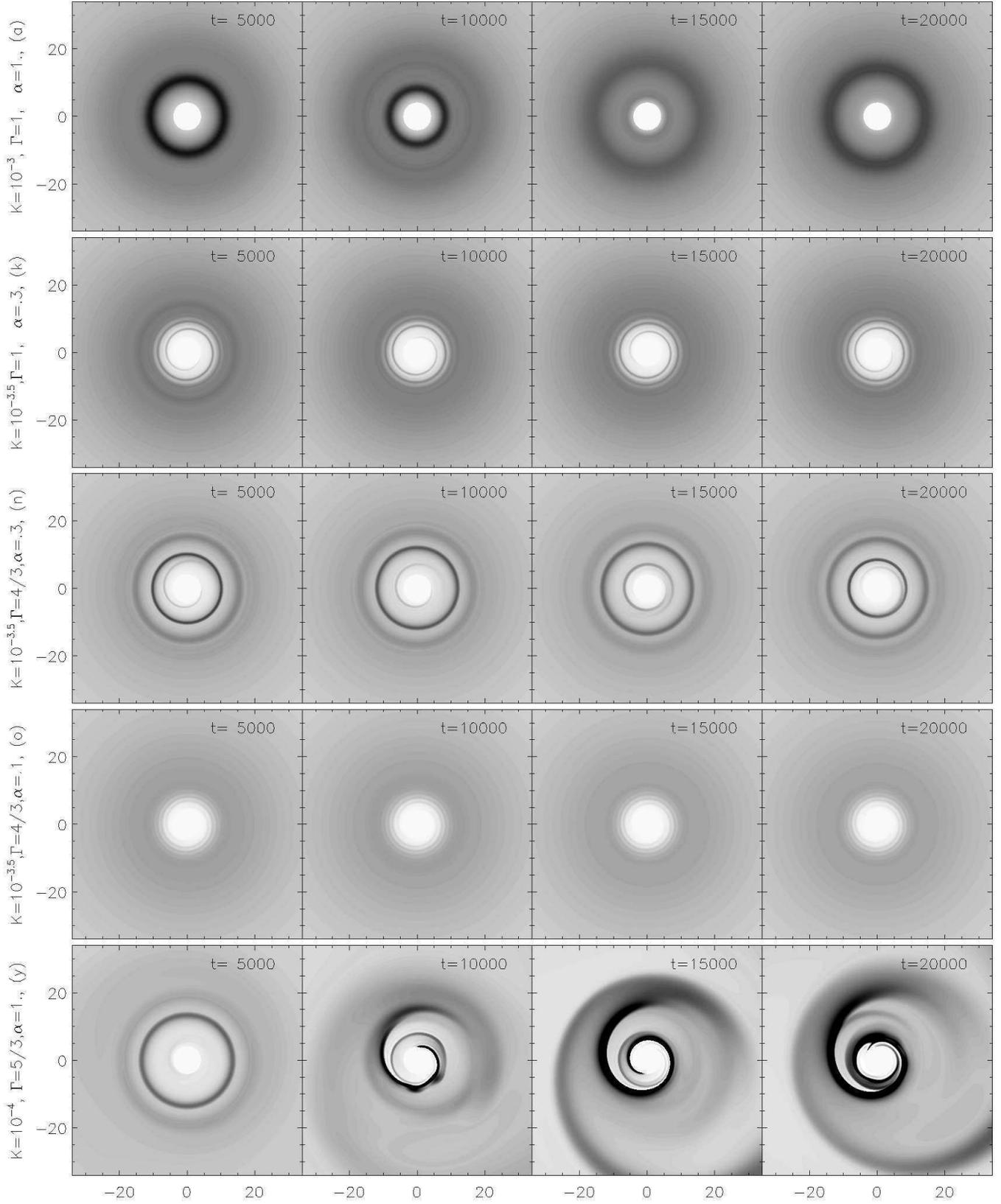}
  \caption{Density contours for a representative set of parameters
    $(K, \Gamma, \alpha)$ at different times.  The gray scale is
    linear, with white representing $\Sigma = 0$ and black
    representing $\Sigma = 6$.  From top to bottom, we have (\emph{i})
    pure axisymmetric waves, (\emph{ii}) axisymmetric waves
    interacting with an one-armed spiral, (\emph{iii}) an one-armed
    spiral in the inner disk, (\emph{iv}) two-armed spirals in the
    inner disk, and finally (\emph{v}) a turbulent state in the inner
    disk, and one-armed spiral in the outer disk.  See
    \S\ref{sec:results} for an overview of these features.}
  \label{fig:contours}
\end{figure*}

Figure~\ref{fig:contours} shows a representative set of snapshots from
our parameter study.  Each row contains four density contours of a
particular simulation.  The gray scale is linear, with white
representing $\Sigma = 0$ and black representing $\Sigma = 6$.  From
top to bottom, we can see the five most significant features:
(\emph{i}) A pure axisymmetric mode is excited near the ISCO and
propagates outward.  (\emph{ii}) An one-armed spiral develops only at
the inner edge, while the other part of the computational domain is
stable.  (\emph{iii}) Axisymmetric rings appear and interact with the
one-armed spiral at the inner edge. (\emph{iv}) Two-armed spirals
develop at the inner edge when the rest of the disk is stable.
Finally, (\emph{v}) the axisymmetric waves clean up the inner part of
the disk.  Inner spirals develop in this low density region and
perturb the rings.  The rings then break apart in late time and go
through a turbulent transition state.  After that, strong steady
trailing spirals develop in the inner disks while global leading
spirals develop in the outer disks.

We will discuss these features in more detail in later sections.
Nevertheless, to provide a general picture, we review the standard
models of thin disk oscillation.  Following
\citetalias{2001PASJ...53....1K}, we consider the small quantity
$h_{mn}(r) \sim P^{(1)} / \Sigma^{(0)}$.  The subscript $m$ and $n$
indicate different azimuthal and vertical modes, respectively.  And
$P^{(1)}$ denotes the first order fluctuation in pressure.  For
simplicity, we will drop the subscripts $m$ and $n$ hereafter.  The
small quantity $h$ is governed by the equation
\begin{equation}
  \cs^2 \frac{d}{dr} \left(\frac{1}{\tilde\omega^2 -
    \kappa^2}\frac{dh}{dr}\right) + \left(1 - \frac{n
    \cs^2}{\tilde\omega^2 H^2}\right)h = 0, \label{eq:h}
\end{equation}
where $\tilde\omega \equiv \omega - m\OK$ and $\kappa$ is the
epicyclic frequency \citep{1992ApJ...393..697N}.

Assuming the perturbation $h$ is local and has radial wavenumber $k$,
the previous equation reduces to a dispersion relation
\begin{equation}
  (\tilde{\omega}^2 - \kappa^2) (\tilde\omega^2 - n \Omega_\perp^2)
  = \tilde\omega^2 \cs^2 k^2,
\end{equation}
where $\Omega_\perp = \cs / H = \OK$ is the vertical oscillation
frequency.  Setting $n = 0$ for a two-dimensional disk, the $p$-mode
is described by
\begin{equation}
  \tilde{\omega}^2 \equiv (\omega - m\OK)^2 =
  \kappa^2 + \cs^2 k^2; \label{eq:p}
\end{equation}
while the $g$-mode reduces to a trivial mode
\begin{equation}
  \tilde{\omega} \equiv \omega - m\OK = 0, \label{eq:g}
\end{equation}
which describes pure rotations.  We will argue in \S\ref{sec:rings}
that the axisymmetric rings seen in cases (\emph{i}) and (\emph{iii})
cannot be described by setting $m=0$ in equation~(\ref{eq:p}).  On the
other hand, the non-axisymmetric spirals in (\emph{ii}), (\emph{iii}),
and (\emph{iv}) can be described by equation~(\ref{eq:g}), while the
global spirals are the $m=1$ $p$-modes.

\LongTables
\begin{deluxetable*}{cccccccccccc}
  \tablecaption{Summary of results from the parameter study}
  \tablehead{$K$ & $\Gamma$ & $\alpha$ & Label & Resolutions &
    $\dot{M}_\mathrm{ana}$ & $\dot{M}_\mathrm{num}$ &
    Init. $\oring$ & $\oring$ & 
    Init. $\ospiral$ & $\ospiral$ & Global Spiral}
  \startdata
    $10^{-3}$   & $1$   & $1$        & (a) & $129\times32$  & 8.603 & 8.390 & $\approx$ 0.02 & 0.015 & & & \\
    $10^{-3}$   & $1$   & $1^{-0.5}$ & (b) & $129\times32$  & 2.720 & 2.525 & $\approx$ 0.03 & 0.028 & & & \\
    $10^{-3}$   & $1$   & $0.1$      & (c) & $129\times32$  & 0.860 & 0.767 & decay          &       & & & \\
    $10^{-3}$   & $4/3$ & $1$        & (d) & $129\times32$  &11.470 &11.223 & $\approx$ 0.01 & 0.012 & & & \\
    $10^{-3}$   & $4/3$ & $1^{-0.5}$ & (e) & $129\times32$  & 3.627 & 3.411 & $\approx$ 0.03 & 0.025 & & & \\
    $10^{-3}$   & $4/3$ & $0.1$      & (f) & $129\times32$  & 1.147 & 1.032 & decay          &       & & & \\
    $10^{-3}$   & $5/3$ & $1$        & (g) & $129\times32$  &14.338 &14.049 & $\approx$ 0.01 & 0.009 & & & \\
    $10^{-3}$   & $5/3$ & $1^{-0.5}$ & (h) & $129\times32$  & 4.534 & 4.306 & $\approx$ 0.02 & 0.025 & & & \\
    $10^{-3}$   & $5/3$ & $0.1$      & (i) & $129\times32$  & 1.434 & 1.297 & decay          &       & & & \\
  \hline                                                                                                                   
    $10^{-3.5}$ & $1$   & $1$        & (j) & $257\times64$  & 2.720 & 2.635 & $\approx$ 0.02 & 0.021 &                      &                & \\
    $10^{-3.5}$ & $1$   & $1^{-0.5}$ & (k) & $257\times64$  & 0.860 & 0.810 & $\approx$ 0.03 & 0.031 & $\approx$ 0.10--0.11 & 0.098          & \\
    $10^{-3.5}$ & $1$   & $0.1$      & (l) & $257\times64$  & 0.272 & 0.250 & decay          &       &                      & 0.212          & \\
    $10^{-3.5}$ & $4/3$ & $1$        & (m) & $257\times64$  & 3.627 & 3.529 & $\approx$ 0.02 & 0.018 &                      &                & \\
    $10^{-3.5}$ & $4/3$ & $1^{-0.5}$ & (n) & $257\times64$  & 1.147 & 1.089 & $\approx$ 0.03 & 0.031 & $\approx$ 0.09--0.12 & $\approx$ 0.01 & \\
    $10^{-3.5}$ & $4/3$ & $0.1$      & (o) & $257\times64$  & 0.363 & 0.338 & decay          &       & $\approx$ 0.21       & 0.206          & \\
    $10^{-3.5}$ & $5/3$ & $1$        & (p) & $257\times64$  & 4.534 & 4.425 & $\approx$ 0.02 & 0.018 &                      &                & \\
    $10^{-3.5}$ & $5/3$ & $1^{-0.5}$ & (q) & $257\times64$  & 1.434 & 1.367 & $\approx$ 0.03 & 0.028 & $\approx$ 0.06--0.15 & $\approx$ 0.01 & \\
    $10^{-3.5}$ & $5/3$ & $0.1$      & (r) & $257\times64$  & 0.453 & 0.425 & decay          &       & $\approx$ 0.20       & 0.202          & \\
  \hline                                                                                                                    
    $10^{-4}$   & $1$   & $1$        & (s) & $513\times128$ & 0.860 & 0.842 & $\approx$ 0.02 & 0.025 & decay          & weak  &        \\
    $10^{-4}$   & $1$   & $1^{-0.5}$ & (t) & $513\times128$ & 0.272 & 0.264 & $\approx$ 0.03 & 0.031 & decay          & weak  &        \\
    $10^{-4}$   & $1$   & $0.1$      & (u) & $513\times128$ & 0.086 & 0.082 & decay          &       &                & 0.117 &        \\
    $10^{-4}$   & $4/3$ & $1$        & (v) & $513\times128$ & 1.147 & 1.049 & $\approx$ 0.02 & ---   & decay          & ---   & steady \\
    $10^{-4}$   & $4/3$ & $1^{-0.5}$ & (w) & $513\times128$ & 0.363 & 0.349 & $\approx$ 0.03 & ---   & decay          & ---   & steady \\
    $10^{-4}$   & $4/3$ & $0.1$      & (x) & $513\times128$ & 0.115 & 0.110 & decay          &       &                & 0.114 &        \\
    $10^{-4}$   & $5/3$ & $1$        & (y) & $513\times128$ & 1.434 & 0.999 & $\approx$ 0.02 & ---   & decay          & ---   & steady \\
    $10^{-4}$   & $5/3$ & $1^{-0.5}$ & (z) & $513\times128$ & 0.453 & 0.229 & $\approx$ 0.03 & ---   & decay          & ---   & steady \\
    $10^{-4}$   & $5/3$ & $0.1$      & (@) & $513\times128$ & 0.143 & 0.138 & decay          &       & $\approx$ 0.10 & 0.110 &    
  \enddata
  \label{tab:para}
\end{deluxetable*}

\begin{figure*}
  \includegraphics[scale=0.75,trim=18 18 0 12]{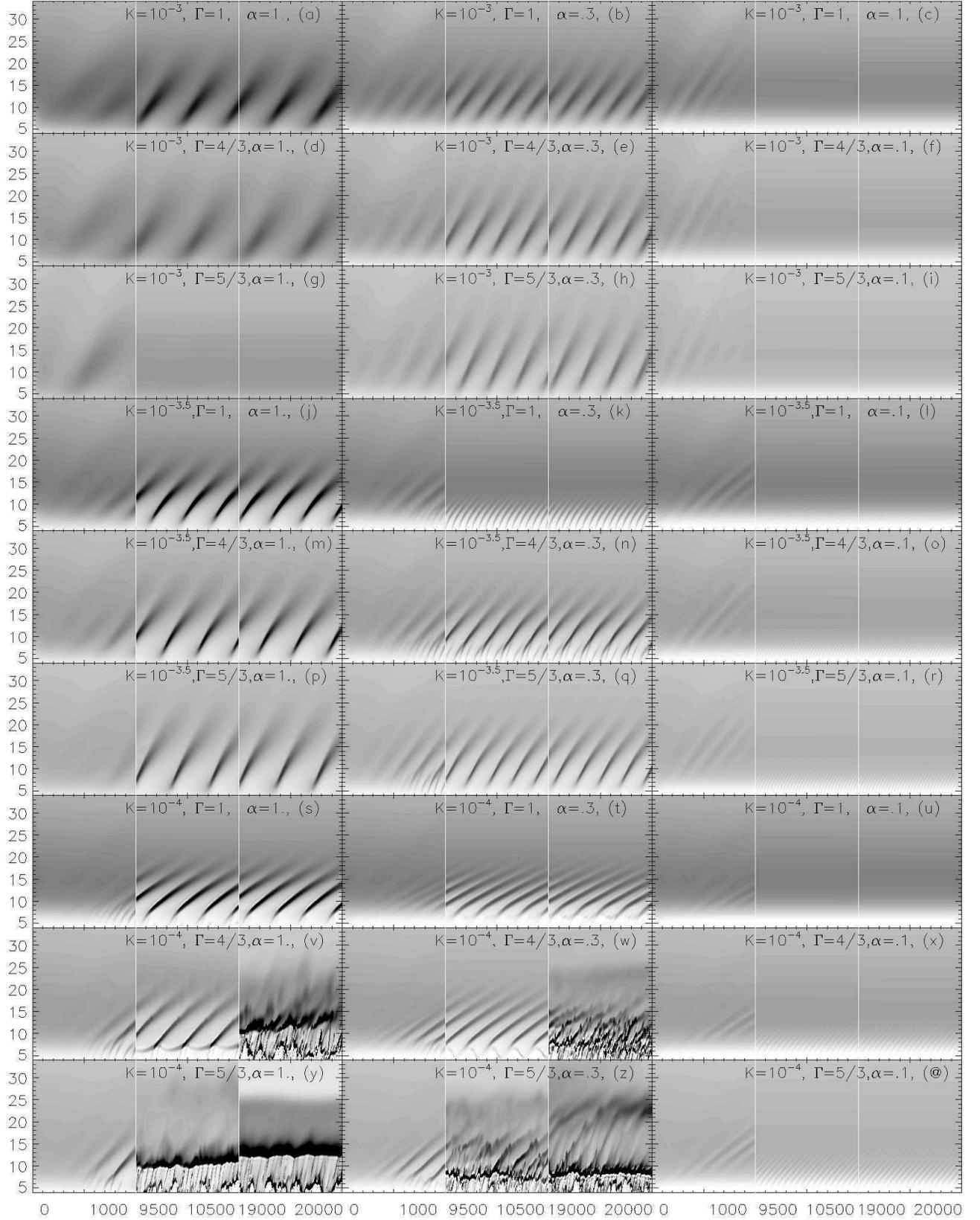}
  \caption{Column density at $\phi = 0$ as contours of $t$ (horizontal
    axis) and $r$ (vertical axis).  The gray scale is linear, with
    white representing 0 and black representing 6.  The horizontal
    axis is chopped into three segments, namely, the early time
    $[0,1000]$, middle time $[9500,10500]$, and late time
    $[19000,20000]$.  The high density (dark) features that appear
    around $r \gtrsim 6$ correspond to the axisymmetric rings.  The
    weaker features that appear around $r \lesssim 12$, with fast
    temporal variability, are non-axisymmetric spirals.}
  \label{fig:histories}
\end{figure*}
\begin{figure*}
  \includegraphics[scale=0.75,trim=18 18 0 12]{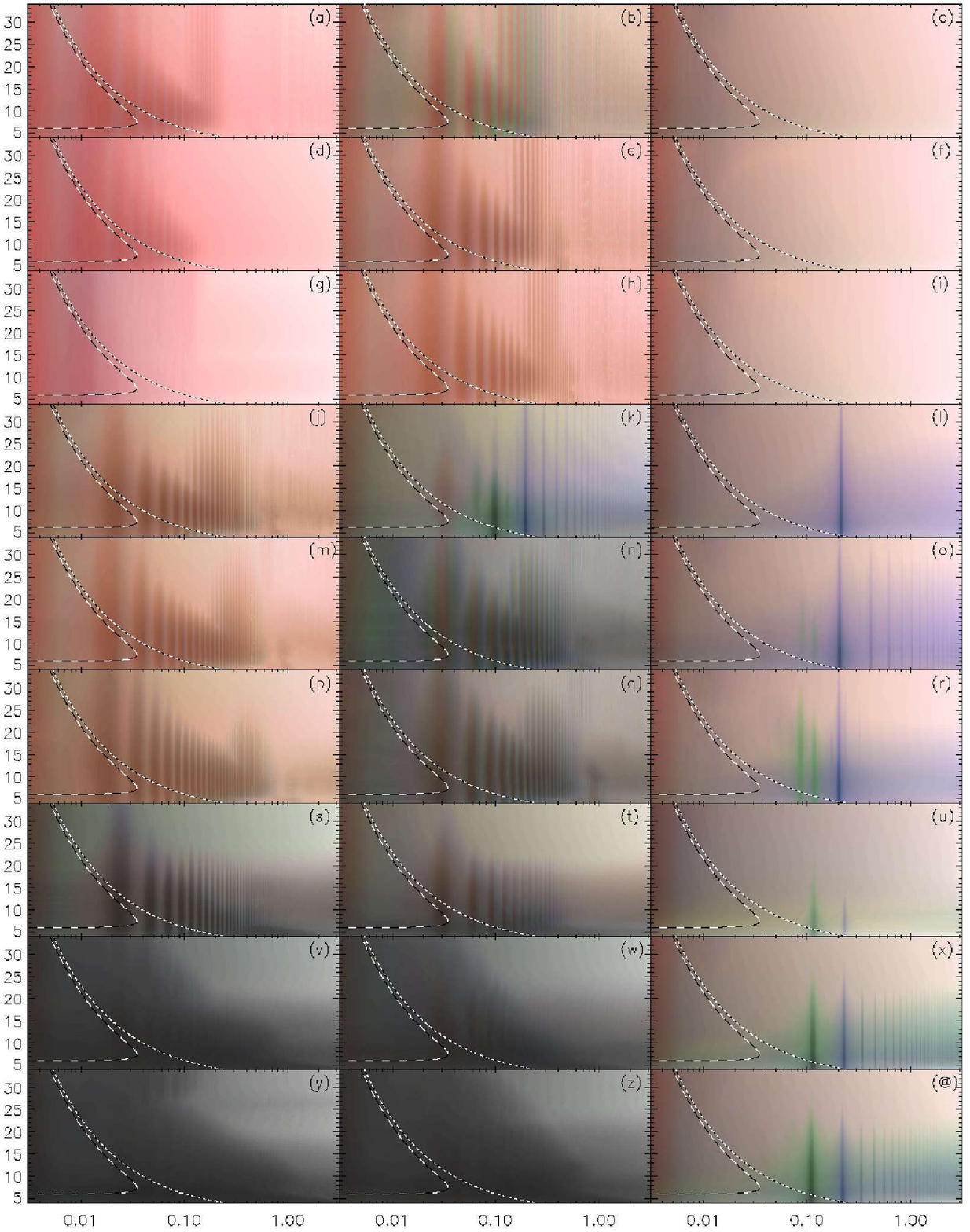}
  \caption{The power spectra multiplied by $\omega$, i.e., $\omega
    P(\omega)$, of the density as contours of frequency (horizontal
    axis) and radius (vertical axis).  They are roughly the Fourier
    transforms of the panels in Figure~\ref{fig:histories} along the
    horizontal axis.  The order of each panel is the same as in
    Figure~\ref{fig:histories}.  Because the first three azimuthal
    modes are the most important ones, we can use red, green, and blue
    to present $m = 0$, 1, 2 modes, respectively.  Deeper color
    represents higher power in the logarithmic scale.  See
    \S\ref{sec:results} and footnote 3 for the details of the coloring
    scheme.  The dotted lines and dashed lines are the Keplerian and
    epicyclic frequencies, respectively.}
  \label{fig:specontours}
\end{figure*}

\vspace*{-12pt}
In order to give a more quantitative description of the various
simulations, we list some important numbers in Table~\ref{tab:para}.
The first three columns are the parameters we explored.  The fourth
column gives a single letter label to each simulation.  The fifth
column lists the resolutions.  Note that, as we decrease $K$, the
kinematic viscosity $\nu_\mathrm{SS}$ decreases as well.  Higher
resolutions are then used to resolve the high Reynolds number
flows\footnote{We have performed very high resolution benchmark
  simulations (not shown in this paper).  These benchmarks are done in
  a smaller computational domain with both constant kinematic
  viscosity and $\alpha$-viscosity.  Both the axisymmetric rings and
  inner spirals appear, implying they are robust features in
  two-dimensional viscous disks.  However, for the cases with extreme
  Reynolds numbers, our computational resource does not allow us to
  run a long enough super-high resolution simulation.  Therefore, it
  is not clear to us whether the turbulent transient state is a
  numerical artifact or a physical phenomenon.  See
  \S\ref{sec:limitations} for detailed discussions.}.  The sixth
column lists the analytical accretion rate $\dot{M}_\mathrm{ana}$
calculated by equation~(\ref{eq:accretion_rate}); while the seventh
column lists the numerical accretion rate $\dot{M}_\mathrm{num}$
computed from the simulations,
\begin{equation}
  \dot{M}_\mathrm{num} = - \frac{1}{T} 
  \int\!\!\!\!\int \Sigma v_r r\;dt\;d\phi.
\end{equation}
The time integrations are taken between $t = 10000$ and 20000 so that
$T = 10000$.  The resulting $\dot{M}_\mathrm{num}$ are almost
independent of the radius.  The values listed in Table~\ref{tab:para}
are computed at $r = 34$, the mid-point of our radial domain, to
minimize the boundary effects.  The differences between
$\dot{M}_\mathrm{ana}$ and $\dot{M}_\mathrm{num}$ are always less than
10\% except for simulations (y) and (z).  This agreement, together
with the fact that the mean density profiles are well described by our
analytical approximations (see next section and
Figure~\ref{fig:profile}), imply that our initial conditions are
reasonably close to the steady states.

The eighth and ninth columns are the oscillation frequencies of the
axisymmetric rings $\oring$.  The values in the table are computed at
the ISCO although the frequencies are inpendent of the radius in all
of our simulations.  The eighth column lists rough descriptions of the
axisymmetric rings in the early time of the simulations.  If the
axisymmetric rings decay away before $t = 2,000$, we identify them as
relaxation effects of the initial conditions and mark ``decay'' in the
table.  If the oscillations remain strong at $t = 2,000$, we measure
the time difference between the density peaks and estimate the
frequency by the equation $\oring = 2\pi/$period.  The values listed
in the ninth column are more robust measurements of the frequencies.
They are calculated based on (temporal) spectral analysis of the
simulations between $t = 7,712$ and 20,000.  The details can be found
in the next section.  The symbol ``---'' indicates the power is
distributed across a wide dynamic range, i.e., no specific frequency
can be identified.  Empty cell indicates the solution is stable so no
significant frequency is found.

The tenth and eleventh columns are the oscillation frequencies of
non-axisymmetric inner spirals $\ospiral$.  Note that, due to the
interaction between inner spirals and the axisymmetric rings, the
spirals do not have a unique frequency in simulations (k), (n), and
(q).  The differences between the peaks are uneven.  We therefore
write a frequency range for the estimated frequencies in the tenth
column.  The eleventh column lists the peak locations for the power
spectra.  Finally, the last column lists if a steady global spiral
appears at late times in the simulations.

\section{Temporal Analysis}
\label{sec:fft}

In order to make the analysis more intuitive, we first plot the column
density for each simulation at $\phi = 0$ as a contour of radius and
time in Figure~\ref{fig:histories}.  The time axis is broken into
three segments, namely, the early time $[0,1000]$, middle time
$[9500,10500]$, and late time $[19000,20000]$.  The gray scale is
linear, with white representing $\Sigma = 0$ and black representing
$\Sigma = 6$.  This is identical to the gray scale used in
Figure~\ref{fig:contours}.

For all panels except (v), (w), (y), and (z), the high density (dark)
features around $r \gtrsim 6$ correspond to the axisymmetric rings.
The slopes of these features, $\delta r/\delta t$, are their (radial)
pattern speeds.  The weaker features appearing around $r \lesssim 12$
correspond to the non-axisymmetric spirals.  Because the angular
velocities are almost Keplerian in all simulations, it is easy to
identify panels (k), (u), (x), and (@) with one-armed spirals, and
panels (l), (o), and (r) with two-armed spirals, simply base on their
temporal variability.

Panels (v), (w), (y), and (z) are more complicated.  The three time
segments show that the properties of the flow are time-dependent.  For
simulations (v) and (w), the flows are laminar in the first two
segments.  The extra features that appear around $r \sim 7$ in the
middle segment indicate that the axisymmetric rings interact with the
very weak inner spirals.  These extra features break apart and transit
to a turbulent state.  The same kind of transition happens in
simulations (y) and (z).  The almost time independent dark feature at
$r \sim 27$ in panel (y) and $r \sim 32$ in panel (z) are the steady
global one-armed spirals (see the lowest row of
Figure~\ref{fig:contours}).

We follow \citeauthor{2001PASJ...53....1K}'s convention and write
\begin{equation}
  \Sigma(t, r, \phi) = \sum_{m} \int_{-\infty}^{\infty}
  \hat\Sigma_m(\omega, r)\;e^{i(\omega t - m\phi)}\;d\omega,
  \label{eq:convention}
\end{equation}
where $\hat\Sigma_m(\omega,r)$ is the Fourier transform/coefficient.
The power spectrum, defined by
\begin{equation}
  P_m(\omega, r) \equiv |\hat\Sigma_m(\omega, r)|^2,
\end{equation}
measures the contribution of different frequencies for a particular
azimuthal mode as function of radius.  Because the dynamic variables
are smooth\footnote{For most of the simulations, the Fourier
  transforms converge exponentially fast along the azimuthal
  direction.  The exceptions are simulations (v), (w), (y), and (z),
  in which the flows go through turbulent transient states.
  Nevertheless, the spectral filters artificially cut off the high
  $m$-modes and make the function smooth at the \emph{grid scale}.}
and periodic along the azimuthal direction, the azimuthal modes are
trivial to obtain by applying discrete Fourier transform.  However,
the variables are not necessary periodic in time, or, at least the
exact period is not known before doing the temporal analysis.  The PDS
obtained by squaring the discrete Fourier transform in a finite time
domain (i.e., periodogram) is not a good estimator of the actual power
spectrum $P_m(\omega, r)$.

In order to improve the statistics, we use the standard ``overlap
method'' \citep[see][]{1992nrca.book.....P} to average over 11
periodograms.  Each periodogram is computed over a section of 2048
snapshots, with two subsequent sections overlapped by 1024 frames.
This requires $12 \times 1024 = 12288$ snapshots in total.  We drop
the first 7712 snapshots in our data to avoid analysing the initial
relaxation.  The Hann window function is applied to each segment
before calculating the periodogram.  The resulting variance in the PDS
is approximately 11\%.

We plot the estimated $\omega P_m(\omega, r)$ in
Figure~\ref{fig:specontours}.  The order of the panels are the same as
Figure~\ref{fig:histories}.  Because only the first few azimuthal
modes are important, we can use the red, green, and blue color
channels to represent each of them.  These three color channels are
overlapped in a way that darker color represents a higher power in a
logarithmic scale\footnote{ Because each color channel is independent
  of each other, we are using the full three-dimensional color space.
  This is very different from the standard one-dimensional color space
  (gray-scale, rainbow, etc).  It is not obvious how to manipulate
  them by using standard vector graphics.  Therefore, in
  Figure~\ref{fig:specontours}, we first ``pixelize'' the gray scale
  contour for each azimuthal mode.  The red, green, and blue channel
  are then filled by the (maximum allowed) value 255, resulting the
  overall red, green, and blue tones, for the $m = 0$, 1, and 2 modes,
  respectively.  To combine these different image, we use the graphic
  ``multiply'' operator.  That is, each channel in the final image is
  a normalized pixel-wise product of the corresponding channels of the
  three contours.}.  For example, the red tone for panel (a), (d), and
(g) indicates they are purely $m = 0$ axisymmetric modes.  The
brownish color for panel (j), (m), and (p) indicates they are
dominated by $m = 0$ modes, but have minor contributions from the $m =
1$ modes.  The dark gray in panel (s), (v), and (y) indicates all $m =
0$, 1, and 2 modes contribute.  The sharp green and blue lines in
panel (l), (o), (r), (u), (x), and (@) correspond to the fundamental
modes and overtones of the one-armed and two-armed spirals,
respectively.

In addition, the dashed lines in Figure~\ref{fig:specontours} are the
epicyclic frequency $\kappa(r)$ and the dotted lines are the Keplerian
frequency $\OK$ in pseudo-Newtonian gravity.  In contrast to the
standard assumption, the oscillations in our simulations are
\emph{not} locally sub-Keplerian.  Instead, the angular frequencies of
both the axisymmetric rings and the non-axisymmetric spirals are
independent of the radius \citep[see][]{2008arXiv0805.0598M}.  Their
values are listed in the ninth and eleventh columns in
Table~\ref{tab:para}.

\section{Axisymmetric Rings}
\label{sec:rings}

If the axisymmetric rings were only seen in our spectral algorithm,
one would worry that they may be due to numerical artifacts of our
code.  The fact that these oscillations were seen in previous studies
of viscous spreading rings \citep{1986MNRAS.220..593P,
  1991MNRAS.249..684O, 1992MNRAS.254..427O, 1995MNRAS.274...61G} as
well as viscous disk models \citep{1992MNRAS.254..427O,
  1992MNRAS.255...51C, 1995ApJ...441..354C, 1996MNRAS.283..919M,
  2008arXiv0805.0598M, 2008arXiv0805.2950R} suggest they are likely to
be physical, at least under the thin disk assumption.  Recently,
\citet{2008arXiv0805.2950R} performed a detailed temporal analysis of
both hydrodynamic and magnetohydrodynamic simulations.  For their
inviscid hydrodynamic simulations, similar axisymmetric oscillations
are seen in the mid-plane, although the authors associate them with
the $g$-modes. We are also aware of the current study of oscillations
in tilted magnetohydrodynamic disks by Henisey \& Blaes (private
communication).

Analytical studies of radial oscillations of axisymmetric modes in
accretion disks have a long history.  \citet{1984ApJ...287..774B}
studied the overstability of axisymmetric oscillations and
\citet{1993ApJ...409..360L} studied axisymmetric wave propagation in
accretion disks.  A detailed analytical study of the generation and
propagation of these waves is beyond the scope of this paper.
However, a simple comparison suggests that more physics is needed in
the standard model, i.e., \citet{2001PASJ...53....1K}, hereafter
\citetalias{2001PASJ...53....1K}, in order to describe wave
propagation in thin disks.

Note that, when $\oring$ is well defined, it is always smaller than
the maximum epicyclic frequency $\kmax \equiv \kappa(\rpeak) \approx
0.0347$, where $\rpeak = 2 (2 + \sqrt{3})$.  This suggests the
axisymmetric rings are generated as inertial waves and then propagate
outward.  Recalling the dispersion relation for axisymmetric inertial
acoustic wave is $\omega^2 = \kappa^2 + \cs^2 k^2$, because $k^2 > 0$
and the disk is thin, $\cs^2 k^2 \ll \Omega_\perp^2 \sim \kappa^2$, it
leads to a constraint on the \emph{local} oscillation frequency
$\omega \gtrsim \kappa$.  Therefore, \citetalias{2001PASJ...53....1K}
argue that a wave with frequency $\oring < \kmax$ cannot propagate in
the region $[r_1, r_2]$, where $r_1$ and $r_2$ are the Lindblad
resonance points satisfying $\oring = \kappa(r)$.  If the wave is
excited far out and propagates inward, its wavelength becomes infinite
at $r_2$ and is reflected back out.  On the other hand, if the wave is
excited near the inner edge of the disk, it will be reflected back and
forth between the inner edge and $r_1$.  This phenomenon is called
\emph{wave trapping}.

Taking simulation (j) as an example, $\oring = 0.021$.  The standard
picture predicts \emph{no} wave can propagate within the region $[r_1,
  r_2] = [6.203, 12.587]$; waves should be trapped between the inner
edge (near the ISCO) and $r_1$.  Panel (j) in
Figure~\ref{fig:specontours} clearly contradicts this prediction and
violates the constraint $\omega \gtrsim \kappa$.  Indeed, the power at
the dominant mode increases in the forbidden region $[r_1, r_2]$, and
extends far to the outer region.  \citetalias{2001PASJ...53....1K}
commented that the physical reasons for the difference between
analytical models and numerical simulations are not clear.

\begin{figure}
  \includegraphics[scale=0.75,trim=18 6 12 12]{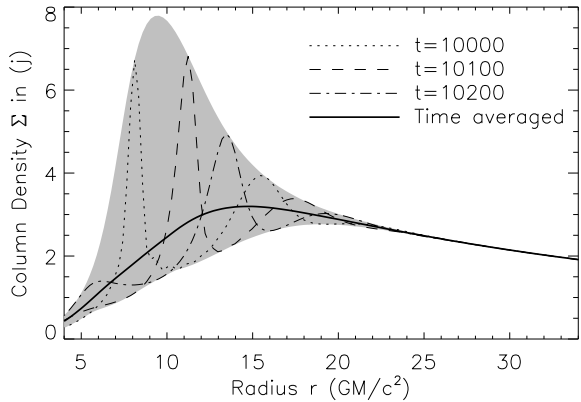}
  \caption{Typical density profiles at $\phi = 0$.  The dotted line,
    dashed line, and dashed-dotted lines are profiles taken at $t =
    10000$, $10100$, and $10200$, respectively.  The thick solid line
    is the time average between $t = 10000$ and $20000$.  The gray
    area marks the maximums and minimums of the waves.}
  \label{fig:profile}
\end{figure}

In Figure~\ref{fig:profile}, we plot the density profile of simulation
(j) at $\phi = 0$.  The different dashed and dotted lines are profiles
taken at $t = 10000$, $10100$, and $10200$.  The thick solid line is
the time average between $t = 10000$ and $20000$.  The gray area marks
the maximums and minimums of the waves.  The amplitudes of the
oscillations (for this particular simulation) are not small compared
to the background.  The peak is located around $r = 9.5$, with peak
densities about four times the mean density.  Also, the wavelengths
are comparable to the radius, i.e, $k \sim 1/r$.

\begin{figure}
  \includegraphics[scale=0.75,trim=18 6 12 12]{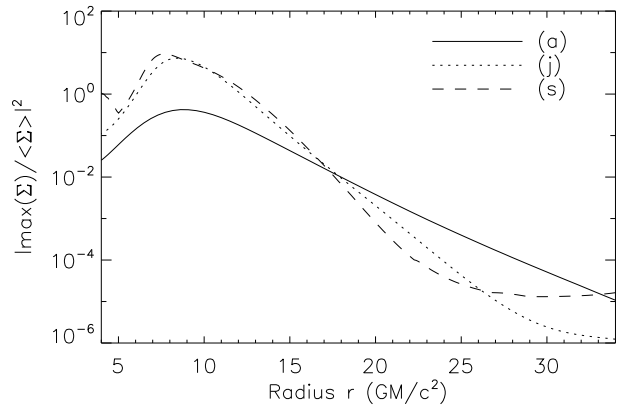}
  \caption{The function $|{\max}(\Sigma)/\langle\Sigma\rangle - 1|^2
    \sim |h|^2$ plotted in log scale taken from the period $t = 10000$
    to 20000.  The solid line is for simulation (a), the dotted line
    for (j), and the dashed line for (s).  When the amplitude is
    small, i.e., for simulation (a), the fluctuation decays
    exponentially for $r \gtrsim 10$.}
  \label{fig:peak2}
\end{figure}

To compare how the amplitudes vary with the parameters, we plot the
quantity $|{\max}(\Sigma)/\langle\Sigma\rangle - 1|^2 \sim |{h}|^2$ in
Figure~\ref{fig:peak2}.  The solid, dotted, and dashed lines
correspond to simulations (a), (j), and (s), respectively.  The
fluctuations in simulation (a) decays exponentially as a function of
$r$.  As the pressure decrease, like in cases (j) and (s), the
fluctuations become nonlinear and the decay rates increase.

There are two kinds of difficulties in applying the standard thin disk
oscillation modes.  First, as shown in Figure~\ref{fig:profile}, the
local linear approximation breaks down due to the non-linearity and
large wavelength of the axisymmetric waves.  Second, even in the cases
where local linear approximation is valid [say simulation (a)],
Figure~\ref{fig:peak2} shows the fluctuations are damped in large
radii, hence the wavenumber $k$ is complex.  The standard approach,
i.e., equation~(\ref{eq:h}), does not take this into account.  More
physics perhaps related to viscosity should be included in the linear
mode analysis in order to understand the wave generation and
propagation.

\section{Non-Axisymmetric Inner Spirals}
\label{sec:spirals}

\begin{figure*}
  \includegraphics[scale=0.75,trim=18 18 0 12]{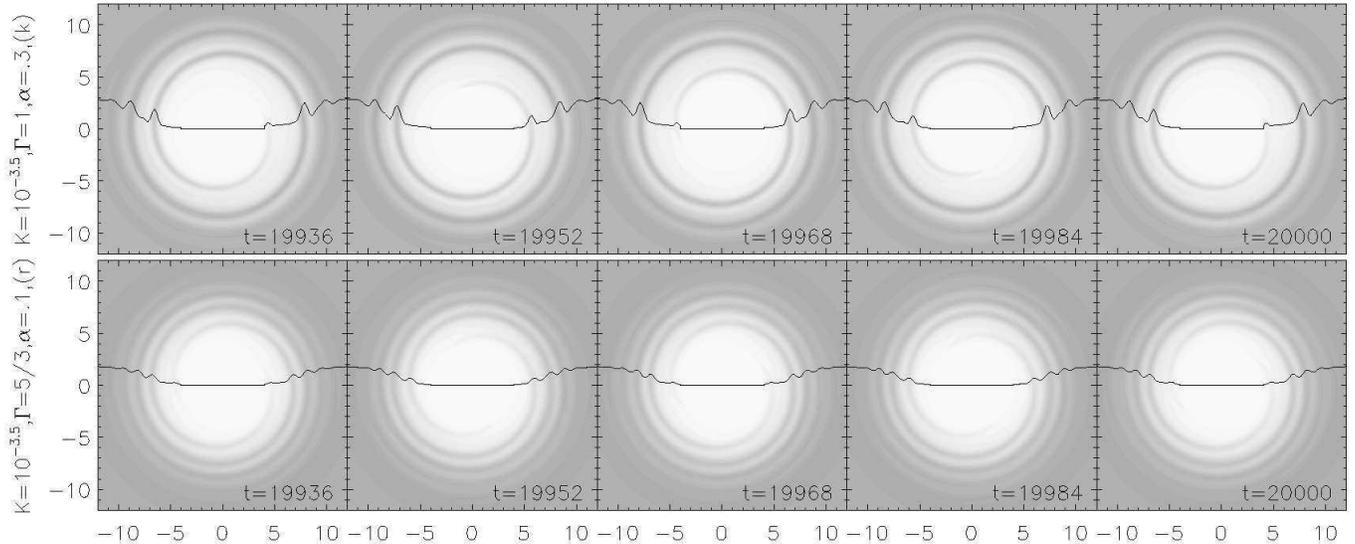}
  \caption{Close up view of the central one-armed (top) and weak
    two-armed (bottom) spirals.  The gray scale is linear in both rows
    with white representing $\Sigma = 0$.  Black represents $\Sigma =
    9$ for the top row and $\Sigma = 6$ for the bottom row.  The time
    difference between each frame is 16, which is about a quarter of a
    period at the ISCO.  Hence, the contours at $t = 19936$ is pretty
    much the same as $t = 20000$ because the flows rotate a full
    period.  The solid lines show the column density along the
    horizontal axis, with the vertical axis matching the numerical
    values.  The solid lines for the bottom row are symmetric because
    of the $m = 2$ modes.}
  \label{fig:weak}
\end{figure*}

Linear stability analysis of hydrodynamic disks was an active area of
study in the 1980's.  \citet{1984MNRAS.208..721P,
  1985MNRAS.213..799P}, \citet{1985MNRAS.213P...7G} and
\citet{1987MNRAS.225..267P} showed that inviscous tori are unstable to
non-axisymmetric perturbations.  The global analysis was extended by
\citet{1985MNRAS.216..553B}, \citet{1986MNRAS.221..339G}, and
\citet{1987MNRAS.225..695G}; while \citet{1987MNRAS.228....1N}
performed a local analysis based on the shearing sheet approximation.
Self-gravity was included in \citet{1988MNRAS.231...97G}, and
\citet{1988ApJ...331..838P} showed that, if the effective viscous
stress is a rapidly increasing function of the column density,
non-axisymmetric perturbations grow because of viscous overstability.

Table~\ref{tab:para} shows that for one-armed spirals, $\ospiral$ is
always very close to the Keplerian frequency at the ISCO, $\OK(\rISCO)
\approx 0.102$; while for two-armed spirals $\ospiral \approx
2\OK(\rISCO)$.  This observation strongly suggests that inner spirals
are generated by \emph{corotating} density fluctuations, which can be
described by the trivial mode at the ISCO.  The density fluctuations
spiral inward as the SPBF suggests and create the rotating patterns in
the inner disks.

In Figure~\ref{fig:weak}, we plot the contours of the very central
region of two simulations that have inner spirals but no significant
axisymmetric rings.  The gray scale is again linear, with white
representing $\Sigma = 0$.  Black represents $\Sigma = 9$ for the top
row [simulation (k) with an one-armed spiral] and $\Sigma = 6$ for the
bottom row [simulation (r) with two-armed spirals].  The time
difference between each snapshot is 16, which is about a quarter of a
period at the ISCO.  Therefore, the contours at $t = 19936$ look
almost identical to the contours at $t = 20000$.

The solid lines in Figure~\ref{fig:weak} show the amplitudes of the
column density $\Sigma$ along the horizontal axis with the values
matching the vertical axis.  The amplitudes are parameter dependent.
For example, the spirals have non-linear amplitudes for simulation
(k); while for simulation (r), the amplitude is small, so perturbation
methods are applicable.  This result is consistent with the analytical
study of \citet{1988ApJ...331..838P}, in the sense that the amplitude
of the spirals is an increasing function of $\Gamma$.  Also note that
the solid lines are symmetric (because of the $m = 2$ mode) for the
bottom row.

\section{Transient and Steady Global Spirals}
\label{sec:global}

\citet{1983PASJ...35..249K} showed that one-armed inertial-acoustic
waves (with $n = 0$), which are just an eccentric deformation of the
disk plane, have frequencies much lower than $\OK$.
\citeauthor{2001MNRAS.325..231O} unified a number of efforts
\citep[see reference in ][]{2001MNRAS.325..231O} and derived a
comprehensive set of evolutionary equations for eccentric disks.
Later, \citet{2003A&A...399..395S} performed a careful perturbation
analysis\footnote{For a small kinematic viscosity, because the
  perturbation is in the highest order term, the problem is singular.
  A stretching time transformation is needed to take into account the
  two different time-scales, as done is \citet{2003A&A...399..395S}.}
and showed that the perturbations are independent of the dynamical
time scales.  These types of (almost) steady global spiral patterns
are believed to be the cause of the V/R variations of Be stars (see
\citetalias{2001PASJ...53....1K} and reference therein).

Following \citet{1983PASJ...35..249K}, we consider the $p$-mode
dispersion relation~(\ref{eq:p}) with $m = 1$,
\begin{equation}
  \omega = \OK \pm \sqrt{\kappa^2 + \cs^2 k^2}.
\end{equation}
For nearly Keplerian disks, $\kappa \approx \OK$.  If the disk is thin
and the oscillation is global, i.e., $\cs k \ll \OK$, the lower frequency
can be approximated by
\begin{equation}
  \omega \approx \OK - \sqrt{\OK^2 + \cs^2 k^2} \approx
  \frac{1}{2}\OK\left(\frac{\cs k}{\OK}\right)^2 \ll \OK.
  \label{eq:steady}
\end{equation}
Hence, for a simulation with a small enough sound speed, we expect our
numerical solution to evolve to a state that agrees with the above
analytical result.

Form the bottom panels of Figure~\ref{fig:contours}, we can briefly
see the different stage of such an evolution.  When $t \lesssim 5000$,
axisymmetric waves are excited and clean up the inner part of the
disk.  Spiral structures can form in this very low density region.
The inner spirals interact with the axisymmetric rings.  At $t \sim
10000$, if the Reynolds numbers are high enough [simulation (v), (w),
  (y), and (z)], both the axisymmetric rings and the inner spirals are
able to break apart.  The inner disks then go through a turbulent
transition state.  Although the turbulent transient states are
potentially very important, we cannot draw strong conclusions from
them.  This is because these high Reynolds number flows are barely
resolved even with $513 \times 128$ grids.  Other concerns include the
inner boundary effects \citep[see][]{2002ApJ...573..728M} and some
numerical issues are discussed in the next section.  Nevertheless, the
strong trailing spirals that develop in the inner disks and the global
leading spirals develop in the outer disks later on are very robust
features.

\begin{figure*}
  \includegraphics[scale=0.75,trim=18 18 0 12]{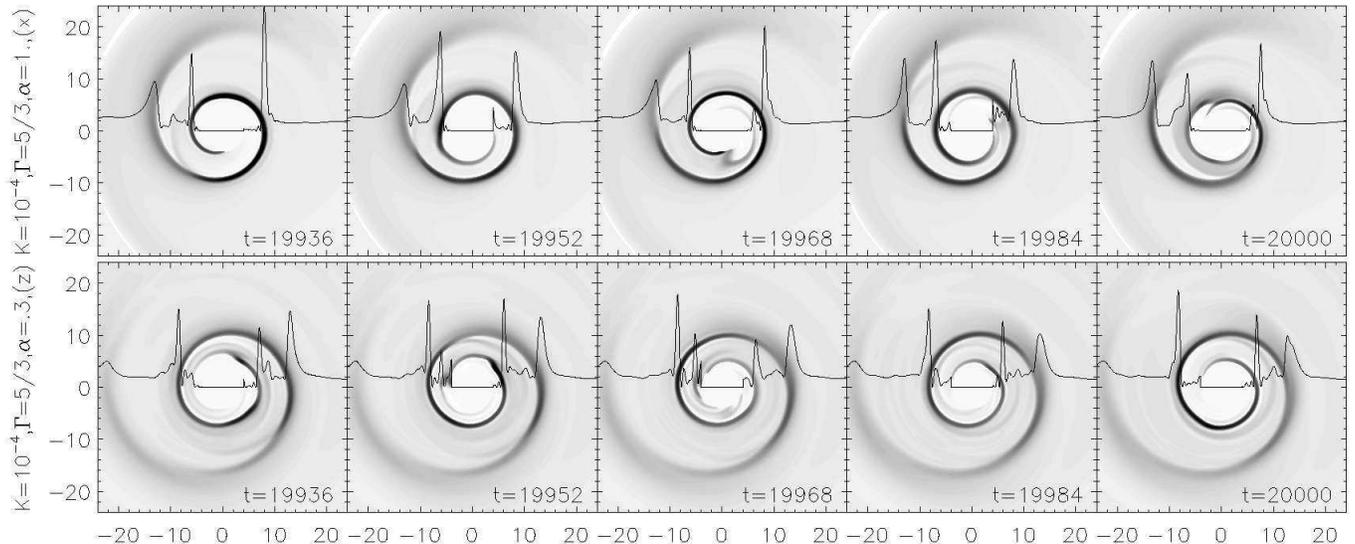}
  \caption{Close up view of the central strong one-armed spirals found
    in both simulation (x) [top row] and (z) [bottom row].  The
    contours are plotted at the same time as in Figure~\ref{fig:weak}
    for easy comparison.  Note that the spirals patterns are almost
    independent of time, as suggested in \citet{1983PASJ...35..249K}.
    The gray scale is again linear, with white representing $\Sigma =
    0$ and black representing $\Sigma = 18$ for both rows.  The solid
    lines show the column density along the horizontal axis, which
    show the spirals are very non-linear.}
  \label{fig:strong}
\end{figure*}

To focus on the trailing spirals, we plot the contours of the central
region of simulation (x) and (z) in Figure~\ref{fig:strong}.  The
contours are plotted using the same time scale as
Figure~\ref{fig:weak} for easy comparison.  The gray scale is again
linear, with white representing $\Sigma = 0$ and black representing
$\Sigma = 18$ for both rows.  The solid lines show the column density
along the horizontal axis, which show the spirals are very non-linear.
Note that the spirals patterns are almost independent of time, as
suggested in \citet{1983PASJ...35..249K} and summarized in
equation~(\ref{eq:steady}).

\section{Limitations}
\label{sec:limitations}

Accretion disks involve different physics at different scales are very
complicated systems.  Although we have presented results based on
careful numerical studies, one needs to be aware of the limitations in
the simulations.  In this section, we provide a list of some important
concerns we have, and suggest several possible improvements for future
studies.

There are several numerical limitations that we need to be aware of.
First, for the high Reynolds number cases, although $513 \times 128$
grid points are used, we can barely resolve the accretion flows.  Our
artificial viscosity may then have important effects on the flows,
even though they are high order.  Second, in some of our simulations,
the axisymmetric rings (together with the inflow, of course) clean up
the inner region of the accretion disks.  The density is so low that a
density floor is often used to prevent negative values.  Because this
only happen in the inner region, it is likely that the artificially
introduced mass would fall towards the central object very quickly.
However, it is not clear to us how important this effect is on the
properties of our numerical solutions.  Third, because we used an
explicit Eulerian algorithm, the time steps are extremely small due to
the fast flow and small grid at the inner boundary.  Our high
resolution simulations use more than a million time steps.  The
culmination of both truncation and round off errors may start
polluting our numerical solutions.

The above concerns suggest that, although the turbulent transient
states found in our high Reynolds number simulations [i.e., (v), (w),
  (y), and (z)] are potentially important, one should not draw strong
conclusions from them.  Better numerical algorithms with higher
resolutions are needed to study them.

In addition to the numerical issues, there are problems associated
with non-general relativistic hydrodynamics.
\citet{2002ApJ...573..728M} performed a careful study of
two-dimensional (in $r$-$z$ plane) viscous hydrodynamic disks.  They
pointed out that numerical solutions in pseudo-Newtonian gravity
depend on the location of the inner boundary.  In our simulations,
because the chosen sound speeds are small, the flows at the inner
boundary are super-sonic.  The boundary effect is therefore not
important.  Nevertheless, a general relativity version of our code
will improve the numerical solutions and lead to more confident
results.  However, there is no unique way to formulate relativistic
hydrodynamics with viscosity.  Methods like using flux-limited
diffusion \citep{2003rnh..book.....W}, or including an explicitly
defined ``finite time propagator'' \citep{2007BrJPh..37.1047K}, are
proposed.

Finally, we have completely neglected the effect of magnetic fields
and adopted the $\alpha$-viscosity prescription.  It is well known
that MHD turbulence in thick disk simulations tend to destroy coherent
structures and wash away periodic signatures.  Moreover,
\citet{2008MNRAS.383..683P} pointed out that magnetorotational
instability (MRI) driven turbulence does not behave like
$\alpha$-viscosity.  It is not clear whether
equation~(\ref{eq:continuity}) and (\ref{eq:navier-stokes}) can
capture the correct dynamics of accretion flows, even within a thin
disk approximation.  A study of oscillations from MHD simulations of
tilded disks [Henisey \& Blaes (private communication)] and thin MHD
disks \citep{2008arXiv0808.2860S} will therefore be very interesting.

\section{Discussions}
\label{sec:discussions}

In this paper, we have performed a parameter study of two-dimensional
viscous accretion disks.  We have found three types of robust
features: the axisymmetric rings, the non-axisymmetric inner spirals,
and the steady global one-armed spirals.  When applying a temporal
analysis of the column density, a large amount of power is contained
in a few modes with frequencies that are independent of the radius.

Although the properties of the axisymmetric rings do not completely
agree with theoretical predictions, the frequencies $\oring$ are
always smaller than the maximum epicyclic frequency $\kmax$.  This
strongly suggests that they are axisymmetric $p$-modes.  By taking a
closer look at the simulations, we find out that the local, linear
assumption breaks down in the inner disk.  Moreover, based on the
results of the simulations, we argue that the standard linear mode
analysis is not enough to understand these rings.  The fact that these
inertial acoustic modes can propagate to large radii with constant
frequencies have important implications --- a frequency found in an
observed oscillation can correspond to a large range of radii in the
accretion disks.  Therefore, one cannot use it to estimate the size of
the emission region \citep[see][for related
  discussions]{2008arXiv0805.0598M}.

The two types of non-axisymmetric modes have very different
properties.  For the inner spirals, their frequencies are always very
close to a small multiple of the Keplerian frequency at the ISCO, $m
\OK(\rISCO)$.  Hence, they simply are rotating patterns originating at
the ISCO and corresponding to the trivial $g$-mode $\tilde\omega = 0$.
The formation of the $m = 1$ and $m = 2$ modes is probably due to
viscous overstability.

Although there are some situations in which both the axisymmetric
rings and the non-axisymmetric spirals co-exist, the connection
between these two features is not entirely clear.  The excitation of
rings usually cleans up the inner disks and prevents the inner spirals
to develop.  However, there are cases in which that the inner spirals
perturb the axisymmetric rings.  Steady global spirals are then
excited.  These steady patterns have properties that agree with the
low frequency global $m = 1$ $p$-modes.  They are predicted as general
features when the disks weakly deviated from Keplerian and have slow
sound speed \citepalias{2001PASJ...53....1K}.  The same feature can
also be understood as \emph{eccentric deformation}.

Modeling if QPOs can roughly be classified into two major schools.
One conjectures that the observed oscillations correspond to the
eigenfrequencies in the accretion disks.  The other proposes that an
inspiral stream of matter is responsible for the modulation.  Our
simulations show, depending on the parameter, both features can be
excited in viscous disks.  Although the two frequencies are associated
with axisymmetric rings and inner spirals are in ratio 1:3 instead of
the observed 2:3, our results provide new insights to understand and
model QPOs.

\acknowledgements

The author would like to thank Ramesh Narayan, Alexander Tchekhovskoy,
Ruth Murray-Clay, and Robert Penna for useful discussions; Dimitrios
Psaltis, Feryal \"Ozel, Martin Pessah, Sukanya Chakrabarti for giving
detailed comments on the manuscript; Gordon Ogilvie for pointing out
important references on hydrodynamic overstability; Ken Henisey and
Omer Blaes for kindly sharing their on-going work on disk
oscillations.  The simulations presented in this paper were performed
on a Beowulf cluster in the Physics Department at University of
Arizona.  The author is currently supported by an ITC fellowship.

\bibliography{astro}

\end{document}